# The Role of Long-lived Excitons in the Dynamics of Strongly Coupled Molecular Polaritons

Bin Liu[1], Vinod M. Menon [1,2*], Matthew Y. Sfeir[2,3*]

[1]Department of Physics, City College of New York, New York, NY 10031, USA

[2]Department of Physics, Graduate Center, City University of New York, New York, NY 10016, USA

[3]Photonics Initiative, Advanced Science Research Center, City University of New York, New York, NY 10031, USA

* Email: vmenon@ccny.cuny.edu
* Email: msfeir@gc.cuny.edu

**Abstract:** The concept of modifying molecular dynamics in strongly coupled exciton-polariton systems is an emerging topic in photonics due to its potential to produce customized chemical systems with tailored photophysical properties. However, before such systems can be realized, it is essential to address the open questions concerning the nature and strength of electronic interactions between exciton-polaritons and localized excited states in chemical system as well as the proper way to measure such interactions. Here, we use transient optical spectroscopy to investigate dynamical interactions between exciton-polaritons, singlet excitons, and triplet excitons in a molecular singlet fission system that is strongly coupled to an optical microcavity. We identify some of the major limitations to modify molecular dynamics in the strong coupling regime. Simultaneous excitation of cavity polaritons and "reservoir" states, defined as dark polaritons and dark excitons (e.g. triplets) from coupled molecules and excitons from uncoupled molecules, always occurs. In addition, slow conversion from reservoir states to cavity polaritons results in minimal changes to the overall population dynamics. Furthermore, we demonstrate how in addition to the usual population dynamics, transient optical measurements on microcavities reveal information pertaining to modification of the exciton-polariton transition energies due to changes in the population of molecular excited states and the exciton-photon coupling conditions. As a consequence of weak interactions between reservoir states and cavity polaritons, judicious design considerations are required to achieve modified chemical dynamics, necessitating the use of molecular systems with long excited-state lifetimes or strong coupling approaches that require a small number of molecules.

## INTRODUCTION

Organic polaritons are the result of strong coupling between molecular excited states and optical resonant modes (e.g., cavity photons and surface plasmons) at room temperature, which yields a hybrid quasiparticle with partial matter and partial photon character.[1–3] For chemical systems, the growing interest in this subject stems from the potential to non-synthetically tune their properties through modification of their light-matter interactions, including the energy, lifetime, and density of states of electronic and vibrational transitions.[4–9] In particular, organic exciton-polaritons (exciton-photon hybrid states) have been extensively studied for optoelectronic applications,[10–12] energy transfer,[13,14] nonlinear optics,[15–17] condensation,[18–20] superfluidity,[21] transistor action,[22] and in organic spin conversion processes.[23–26] However, molecular systems contain a diverse set of accessible excitonic states with distinct orbital and spin angular momenta, frequently resulting in complex excited state dynamics. As such, the dynamical interactions between exciton-polaritons and other excited states needs to be clearly established before the full realization of tunable strongly coupled chemical systems is achieved.

Organic spin materials, in which dynamic interconversion between singlet and triplet excitons occurs, represent a rich platform to study the influence of strong coupling on the excitonic dynamics of molecular materials. Initial studies integrating these materials into optical cavities have led to controversies about the effect of strong-coupling on kinetic processes, including interactions with optically "dark" exciton states (e.g., triplets and triplet pairs), lifetimes of the polaritons themselves, and the strength of coupling between exciton-polaritons and other excitations in the system.[23–26] For example, two recent experiments examining the effect of cavity polaritons on the reverse intersystem crossing process (energetically allowed conversion from triplet to singlet excitons) reach dramatically different conclusions regarding the conversion kinetics of triplet excitons to cavity polaritons.[23,24] A central issue in this controversy is the presence of a large reservoir of localized "dark" exciton-polariton states (of order $N$-1, where $N$ is the number of molecules coupled to the cavity) that have poor spatial coupling to small number (2) delocalized polariton states.[27] Furthermore, there exist a sub-ensemble of molecules that are uncoupled to the cavity due to misaligned transition dipoles which can affect the relaxation dynamics.

Similarly, several recent reports have considered strong coupling using singlet exciton fission materials (SF),[25,26] in which molecular singlet excitons rapidly convert into a triplet pair (biexciton) state followed by dephasing into two independent triplet excitons.[28] Optoelectronic devices fabricated from singlet fission molecules have exceeded 100% external quantum efficiency,[29,30] thus singlet fission molecules are of growing interest in both fundamental investigations and applications for multi-excitonic devices.[31] These materials are notable for the diversity of their available excited states in terms of the electronic character (exciton/biexciton), spin character (singlet, triplet, and quintet multiplicities), and characteristic time scales for recombination (femto- to microseconds). In the limit where exciton-polaritons and excitonic states can strongly interact, the formation of exciton-polaritons has the potential to drastically alter the dynamics of these systems. This is especially true for singlet fission systems in which the energy of the singlet exciton and biexciton are nearly degenerate, such as tetracene, such that the formation of exciton-polaritons can switch the sign of the energetic driving force (singlet-triplet pair energy difference) for singlet fission subsequently affect the rate of triplet pair formation.[25,26]

However, before the energetics of a singlet fission microcavity system can be systematically controlled, fundamental scientific questions need to be addressed, such as whether exciton-polaritons can directly decay to biexcitonic states, the role of reservoir states (including "dark" polaritons, excitons of higher spin multiplicity, and uncoupled molecules) in determining the relaxation dynamics, and the effect of excited state occupation on the strong coupling conditions. Calculations have suggested that modification of the singlet exciton dynamics on picosecond timescales will occur due to the presence of the polariton states, affecting the triplet pair formation time.[32] In contrast, experiments have mostly indicated relatively weak perturbation of the system, with the most prominent consequence being enhancement of the delayed fluorescence lifetimes and yields, corresponding to modified triplet pair recombination on nano- to microsecond timescales.[25,26] However, only time-resolved photoluminescence techniques have been used to probe strongly coupled singlet fission materials, which are not sensitive to the early time dynamics relevant to generation of triplet pairs (~ 1 ps) or non-emissive excited states. Although transient optical spectroscopy is a more desirable technique since it provides high time resolution (~ 100 fs) and allows for direct identification of emissive and non-emissive transient species via their spectral response,[33–37] surprisingly few examples exist where transient optical methods are directly applied to organic microcavity systems.[38–40] Furthermore, some uncertainty persists about the interpretations of the observed transient features, particularly their relationship to the intrinsic lifetime of the exciton-polariton. As such, the role of reservoir states in the dynamics of strongly coupled molecular polaritons remains an open question.

Here, we experimentally investigate the interconversion of excitons polaritons, singlet excitons and triplet excitons in a singlet fission material strongly coupled to an optical microcavity using transient optical spectroscopy. Despite prominent transient spectral features observed near the polariton resonances, the overall carrier dynamics are nearly identical between the uncoupled and strongly coupled singlet fission system due to slow conversion from reservoir states to exciton-polaritons. Instead of indicating a polariton population, the prominent transient features near the cavity polariton resonances are attributed to changes in the population dynamics of the *reservoir* states, which in turn modify the exciton-photon coupling conditions. As such, the transient response near the cavity polaritons persists even in the absence of any polariton population, extends orders of magnitude longer than their intrinsic lifetime (< 50 fs), and evolves over time. For example, spectral changes follow the conversion of singlet excitons into triplet excitons on timescales of ~ 50 ps followed by decay of triplet excitons on time scales exceeding 10 μs. These observed dynamics can be modeled considering only the population dynamics of the neat film and transient changes in the permittivity (refractive index) of the material resulting from its excited state population. The evolving spectral dynamics demonstrate the extraordinary sensitivity of the strong coupling conditions and Rabi splitting to small changes in the molecular excited state population, even in the limit of weak polariton - reservoir interactions.

**RESULTS AND DISCUSSION**

The singlet fission molecule employed is triisopropysilylacetylene pentacene (TIPS-Pc), with the molecular chemical structure shown in the inset of **Figure 1a**. The TIPS functional group renders the pentacene chromophore soluble and stable, resulting in facile and scalable synthesis.[41] In molecular crystals of TIPS-Pc, intermolecular electronic coupling is significant, giving rise to a large red-shift and broadening of the optical transitions, and increased complexity of the molecular exciton states due to Davydov splitting.[42,43] In order to obtain a large number of oscillators in our microcavity but prevent this large shift and broadening, we prepare thin films of singlet fission molecules dispersed in a transparent polymeric matrix (polystyrene, PS) at a concentration of 30% by mass. We designed our cavity to achieve strong coupling with the lowest energy exciton state ($S_1$) near 650 nm (1.91 eV), resulting in an optimal film thickness of ~ 160 nm that corresponds to a total optical density of ~ 0.2 (**Figure 1a**). The photoluminescence of the neat film (on glass) is typical of organic chromophores, with a small Stokes shift for the maximum at 655 nm and the characteristic set of vibronic peaks forming a mirror image of the absorption (**Figure 1a**). The energetics of the strongly coupled system are designed so that the exciton-polariton states are exoergic for singlet fission, similar to the uncoupled singlet exciton. This arrangement permits us to observe modifications of the energy and spin conversion processes due to the presence of the exciton-polariton without consideration of modifying the overall system energetics.

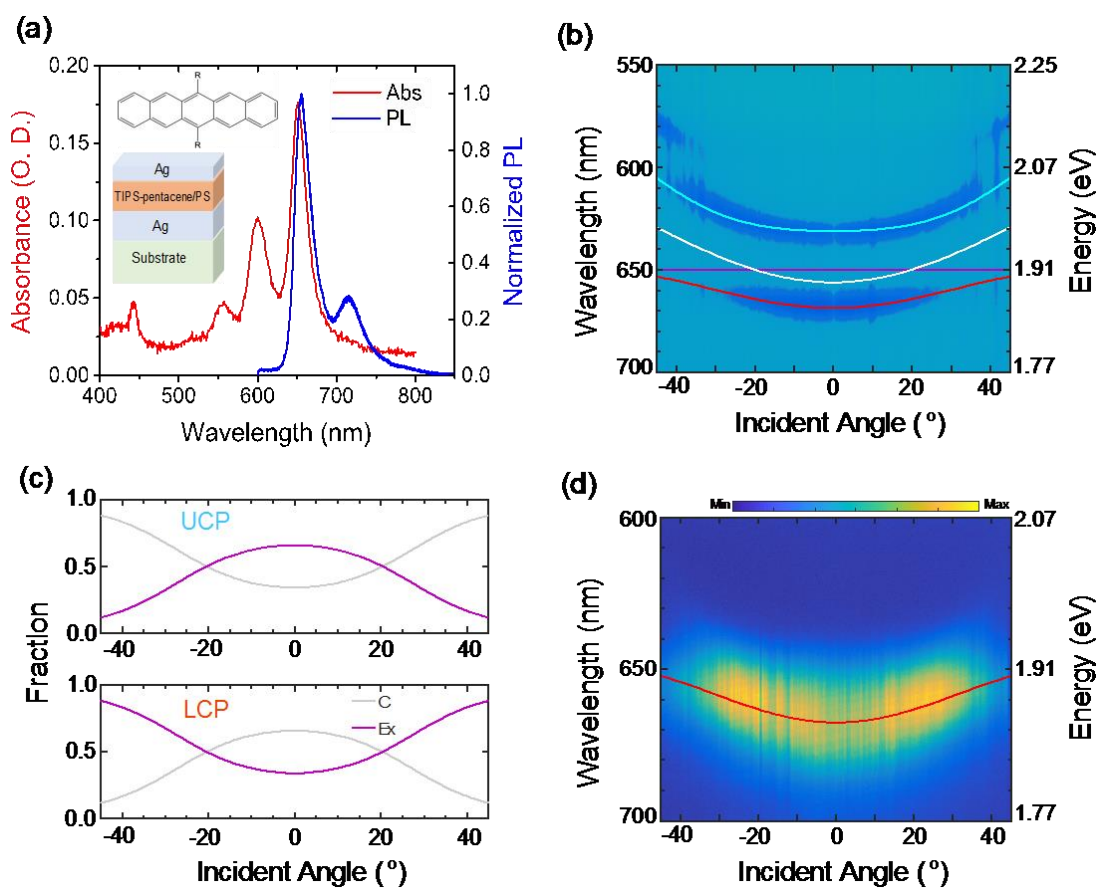

**Figure 1**. a) The absorption (red) and photoluminescence spectrum (blue) of a 160 nm thick film of TIPS-pentacene molecules dispersed into a polystyrene matrix. Inset: the chemical structure of TIPS-pentacene and microcavity structure. b) TE-polarized angle-resolved reflectivity map from a metal cavity containing a 160 nm TIPS-pentacene/PS film. The solid purple and white curves show the uncoupled exciton and cavity photon dispersion; the solid cyan and red curves trace the calculated polariton dispersion using a coupled oscillator model. c) Hopfield coefficients $|\alpha|^2$ and $|\beta|^2$ showing the compositions of UP and LP, where the gray and purple curve represent the fraction of cavity photon and exciton component, respectively. d) TE-polarized angle-resolved photoluminescence map, and the solid red curve indicate the lower polariton dispersion.

To achieve and verify strong coupling in our TIPS-Pc films, we fabricate a planar microcavity and characterize the momentum-energy dispersion of cavity polaritons by angle-resolved reflectivity and photoluminescence (PL) measurements using a Fourier space imaging setup at room temperature.[44] The 160 thick organic film is placed between two metal mirrors to create a Fabry-Perot resonator (inset of **Figure 1a**). The bottom mirror is optically thick (100 nm) and highly reflective, while the top mirror (air interface) is 30 nm thick and partially transparent. This geometry allows optical probing through the top partial reflector. We match the cavity linewidth to the absorption linewidth, resulting in a cavity quality factor (Q) of ~30, corresponding to a cavity lifetime of shorter than 20 femtoseconds (**Section 1** in **Supporting Information**). The contour map of TE-polarized angle-resolved reflectivity reveals the anti-crossing dispersion of upper and lower polaritons (UP, LP) in the strong exciton-photon coupling regime, and two polariton branches anti-cross at an angle of 20° (**Figure 1b**). The polariton dispersion extracted from the reflectivity minima can be fit by a coupled harmonic oscillator model given by[45]

$$\begin{pmatrix} E_c & V \\ V & E_{Ex} \end{pmatrix} \begin{pmatrix} \alpha \\ \beta \end{pmatrix} = E \begin{pmatrix} \alpha \\ \beta \end{pmatrix} \quad (1)$$

where $E_C$ is the cavity mode energy, and $E_{Ex}$ is the uncoupled exciton energy (fixed at 1.91 eV), $V$ is the coupling strength of cavity photons to excitons, and $|\alpha|^2$ and $|\beta|^2$ are the Hopfield coefficients (**Figure 1c**) representing the fraction contribution of cavity photon and exciton component,

respectively. *E* is the angle-dependent cavity polariton eigenenergy, which are traced by the solid cyan and red curves in **Figure 1b**. The cavity photon dispersion is approximated by $E_C(\theta) = E_0(1-\sin^2\theta/n_{eff}^2)^{-1/2}$, where $E_0$ is the cavity cut-off energy (at normal), and $n_{eff}$ is the effective refractive index of metal cavity. The fitting parameter of the coupling strength, *V*, is (55 ± 5) meV, yielding a Rabi splitting of $\hbar\Omega_R = 2V = (110 ± 10)$ meV. It is worth noting that the vibronic overtone of the singlet exciton at 600 nm also strongly couples with cavity photons at large angles (> 40°). The TE-polarized angle-resolved PL, as shown in **Figure 1d**, follows the dispersion of lower polaritons, indicating the dominant radiative relaxation channel occurs through the LP.

Using transient absorption spectroscopy, we verify that singlet fission occurs in our neat films with high yields, producing a diverse set of molecular excited states including singlet excitons, triplet pairs (biexcitons), and triplet excitons. The singlet fission process in TIPS-Pc has been extensively studied in films, micelles, and in individual molecules.[46–51] As such, the spectral signatures of the singlet and triplet states are firmly established, and a convenient spectral fingerprint region for formation of the triplet exists near 508 nm, far from the UP and LP transitions. Here, our control sample consists of a *"half cavity"* structure, with a neat TIPS-Pc/PS film coated on 100 nm Ag high reflector. In our transient measurements, we collect the probe beam in a reflective geometry from the half cavity. Due to the high reflection of bottom silver, changes in the (two-pass) absorption of the molecules are determined by the change in reflectivity $(-\frac{\Delta R}{R})$.

For the *half cavity* system, we see distinct populations of singlets and triplet states that interconvert on rapid time scales similar to all previous work on SF in TIPS-Pc (**Figure 2a**).[46–51] Due to morphological inhomogeneity in our dispersed film, we observe on average, two distinct ensembles of singlets, that decay with distinct characteristic timescales of ~ 5 ps ($S_1^A$) and ~ 60 ps ($S_1^B$). Similarly, triplets recombine on multiple time scales, from tens of nanoseconds to microseconds, suggesting that each singlet sub-ensemble gives rise to a corresponding triplet ensemble ($T_1^A$, $T_1^B$) with its own characteristic decay time (**Figure 2c, left**). For simplicity, we fit the experimental data using a sequential decay scheme (**Figure 2c**, **right**), which allows us to construct characteristic spectra associated with distinct "species" (evolution associated spectra or EAS)[52] that evolve as a function of time according to the equation: $\mathbf{1} \xrightarrow{k_1} \mathbf{2} \xrightarrow{k_2} \mathbf{3} \xrightarrow{k_3} \mathbf{4} \xrightarrow{k_4} \mathbf{S_0}$, with $i$=1-4 distinct spectral decompositions evolving with $k_i$ rate constants and $\mathbf{S_0}$ being the ground state of the system. Importantly, the spectral decompositions allow us to associate observed transient changes with distinct electronic configurations and their characteristic excited state transitions, such as the triplet excited state absorption near 508 nm (**Figure 2b**). Using this approach, we assign the distinct species in the ensemble to linear combinations of individual excitonic species, such that $\mathbf{1} = S_1^A, S_1^B$ (no triplet signal), $\mathbf{2} = T_1^A, S_1^B$, $\mathbf{3} = T_1^A, T_1^B$ (maximum triplet signal), and $\mathbf{4} = T_1^B$ (**Figure 2c**, **right**). The rate constants are determined to be with $k_1$ = 2.0 x 10-1 ps-1, $k_2$ = 1.7 x 10-2 ps-1, and $k_3$ = 1.4 x 10-3 ps-1. The final longer time constant ($k_4$) is not resolved in femtosecond transient absorption measurements but is determined from additional nanosecond transient absorption measurements (**Section 4** in **SI**). We find that there exists a population of triplets that persists for > 10 μs, a lifetime that is similar to triplets in isolated molecules (**Figure S5a**). The long lifetime results from the highly discontinuous TIPS-Pc/PS mixture that results in highly localized (trapped) triplet excitons.

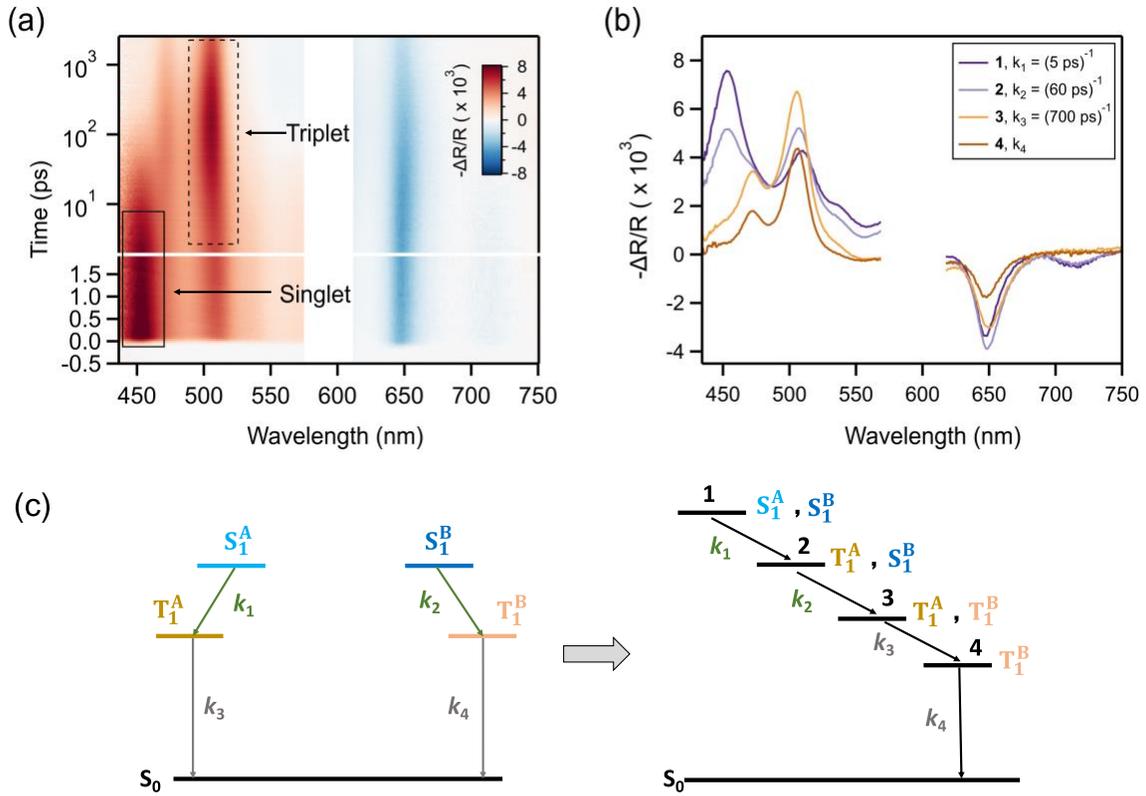

**Figure 2**. a) Transient absorption data of a neat TIPS-Pc/PS film coated on 100nm Ag (*half cavity*), pumped at 600 nm with a fluence of 10 μJ/cm2. Due to pump wavelength scatter, small portion of data have been excluded for clarity. b) Evolution associated spectra (EAS) determined from global analysis of a) using a four species sequential decay model. c) Illustration of the multilevel kinetic scheme for the *half cavity*. The left illustration includes two distinct exciton populations ($S_1^A$, $S_1^B$) that decay independently via singlet fission to triplet excitons ($T_1^A$, $T_1^B$), followed by repopulation of the ground state ($S_0$). The right illustrates the sequential evolution in our simplified model, in which the individual species are a linear combination of the instantaneous exciton population ($S_1^A$, $S_1^B$, $T_1^A$, $T_1^B$).

We find qualitatively similar time-scales for the kinetics in our strongly coupled cavity system, even though the transient spectra change dramatically in the regions near the UP and LP. To quantify the strong coupling conditions in the pump-probe geometry, we first measure the reflectivity spectrum using the probe beam at a small incident angle with the pump beam blocked.

Again, strong dips in the reflectivity spectra indicate the position of the UP and LP (**Figure S2**). Importantly, we observe no significant changes in the dynamics for pumping the exciton reservoir (600 nm), the UP at 630 nm (**Figure S4**), or the LP at 665 nm (**Figure 3a**). The transient absorption spectra of the full cavity system shows several notable features, including a rise near the triplet fingerprint region wavelength (508 nm) and set of prominent dispersive spectral features corresponding to signatures of the cavity polaritons (620-680 nm) indicating strong coupling. We can apply the same sequential model employed to interpret the data on the *half cavity* sample using the same number (4) of excited state species. The resulting fit satisfactorily reproduces the raw data (**Figure 3b**), while the extracted rate constants are nearly identical to those obtained for the half cavity sample, with $k_1 = 3.4 \times 10^{-1}$ ps$^{-1}$, $k_2 = 2.5 \times 10^{-2}$ ps$^{-1}$, and $k_3 = 1.3 \times 10^{-3}$ ps$^{-1}$. As in the *half cavity*, $k_4$ is undetermined in this fit; the long-time scale decay is separately determined using nanosecond transient absorption and again exhibit signals that extends past 10 μs (**Figure S5**, **S6** and **Table S1**). The corresponding spectral decompositions are dominated by signals near the UP and LP transitions, which persist over very long times, comparable to the lifetime of molecular triplet states. Importantly, the decompositions show that the relative amplitude of the UP and LP evolve over time, with time constants that reflect the dynamics of the undressed molecules. Similarly, the feature corresponding to the triplet excited state absorption is readily observable and rises with the expected time constant for direct decay of singlet excitons to triplet pairs (~ 50 ps), confirming that the microcavity system undergoes singlet fission.

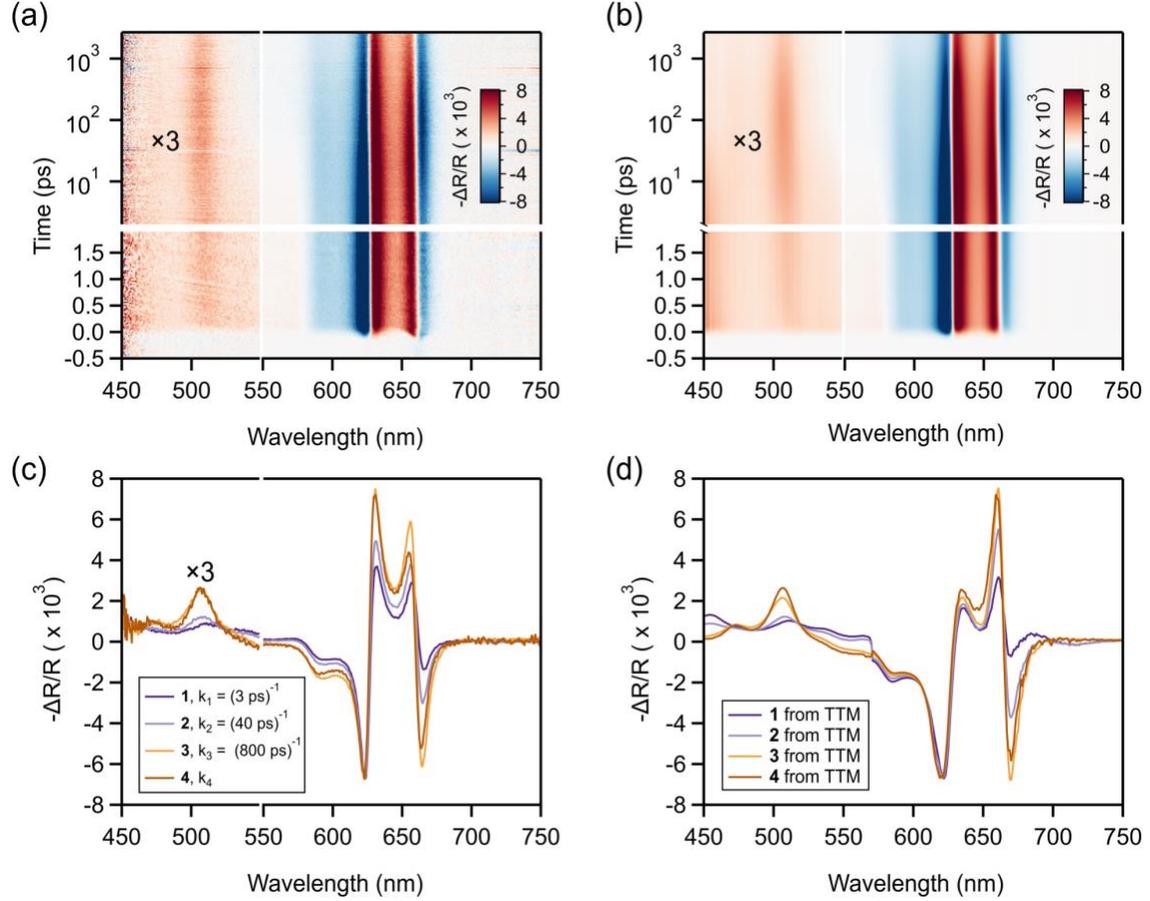

**Figure 3**. a) Transient absorption data of strongly coupled cavity with singlet fission molecules, excited with the LP wavelength of 665 nm with a fluence of 25 μJ/cm2. The data for the wavelength range from 450 nm to 550 nm are increased by 3 times to better represent the kinetics of triplet states. b) Simulated transient absorption using a four exponential global fit of the raw data. c) Evolution associated spectra (EAS) from a sequential decay model involving four species and a set of decay constants connecting them: $\mathbf{1} \xrightarrow{k_1} \mathbf{2} \xrightarrow{k_2} \mathbf{3} \xrightarrow{k_3} \mathbf{4} \xrightarrow{k_4} \mathbf{S_0}$. We note that $k_4$ is too short to be resolved in this data set. d) Calculated transient absorption spectra using the EAS from the *half cavity* (**Figure 2b**) by the transfer matrix method.

To interpret these dynamics, we must develop an effective model that accounts for all the factors that influence the transient optical response of our microcavity system. We can first ignore any dynamics arising from direct population of the LP by the pump pulse. As the polariton has a very short lifetime (< 50 fs) in our low Q cavity system, we are not able to resolve their direct decay given the ~ 100 fs time resolution of our experimental set up. As such, any change observed here in the transient spectra must result from the longer-lived reservoir excitations that are directly populated upon photoexcitation of the system, even when the pump pulse is resonantly tuned to the UP or LP resonance (**Figure S4**). The physics responsible for these long-lived changes in the optical response of system near the UP and LP resonance features comes from two factors: 1) the change of number of coupled molecules, and 2) the modification of effective refractive index which can tune the cavity resonance. We can consider these changes sequentially and estimate the magnitude of their effect on the either the photon dispersion ($\Delta E_c$) or the exciton-photon coupling strength ($\Delta V$) within the coupled harmonic oscillator model (**Equation 1**).

Transient changes in the number of molecules ($\Delta N$) coupled to the cavity will modify the Rabi splitting of the cavity polaritons via changes in the exciton-photon coupling strength: $\Delta V(\Delta N)$. If the excited state absorption spectrum of the molecules is detuned from the cavity dispersion, then photoexcitation will "bleach" a subset of molecules in the ensemble and the number of effective oscillators will be reduced by $\Delta N = I_0(1 - 10^{-A})$, where $I_0$ is the incident number of photons and $A$ is the absorbance. This phenomenon is analogous to the phase space filling or Pauli blocking effect in semiconductor systems, and has been observed in quantum well polariton systems close to saturation.[53–55] As transient absorption is a differential measurement, differences in Rabi splitting in the "pump on" and "pump off" conditions, will give rise to a dispersive response near the LP and UP transitions, even in the absence of a real polariton population. The expected change in the Rabi splitting is estimated to be about 1% under the experimental conditions used in this study (**Section 5** in **SI**). While this is relatively large response, it is not readily observable in our measurement since typical transient optical detection schemes are optimized for high dynamic range but not high spectral resolution (~ 2 nm in our system). As such, the expected shifts of ~ 0.35 nm are not readily observable and do not account for our observed transient dynamics, which show a constant Rabi splitting as a function of both time and input fluence (**Figure S7**). We note that this effect will be enhanced in plasmonic or open cavity systems, due to the large reduction in the number of coupled oscillators.56

Instead, we find that the most important consideration for understanding the spectral dynamics of the full cavity is the change in the effective refractive index of our films after photoexcitation of the molecules. Instead of affecting the Rabi splitting, modification of $\Delta n_{eff}$ affects the photon dispersion $\Delta E_c(\Delta n_{eff})$ and the photon-exciton detuning. The high resolution of our measurement to changes in absorbance (~ 10-4) gives high sensitivity to small modifications to the effective refractive index ($\Delta n_{eff}$), even at low excitation fluences in which a small subset of molecules (< 1%) are photoexcited. Furthermore, the temporal evolution of $\Delta n_{eff}$ after photoexcitation will follow the exciton population dynamics, since both the polarizability of molecules and their absorption spectrum will be distinct for different exciton species (e.g., singlet versus triplet).

The direct relationship between the effective refractive index of our film and the photon-exciton detuning allows us to readily estimate the contribution of the exciton population dynamics to the full cavity transient signal using the spectral decomposition of the *half cavity films* (**Figure 2b**). Briefly, the EAS spectra of the *half cavity* are used to obtain the change in the imaginary part ($\Delta k$) of refractive index and the corresponding modified imaginary ($k' = k+\Delta k$) and real part of refractive index ($n'$) using Kramers-Kronig relations. The complex refractive index for the full cavity system ($n', k'$) are used to calculate the reflectivity in the ground ($R$) and excited state for each distinct exciton populations ($R_i', i = 1-4$) using transfer matrix method as well as the differential transient absorption (($R - R_i'$)/$R$).57,58 The resulting calculated EAS spectra (**Figure**

**3d**) can be compared to those obtained from direct fitting of the full cavity transient absorption with sequential fitting model (**Figure 3c**). Within our model for $\Delta n_{eff}$, the calculated EAS spectra reproduce the dominant spectral changes in the full cavity pump-probe data, capturing the overall signal amplitudes (~ $10^{-3}$), changes in the relative amplitude of the LP and UP bleach features over time, and the rise of the triplet excited state absorption signal. Our model also predicts that the features associated with LP and UP will persist for as long as there is a molecular excited state population. Indeed, using nanosecond transient absorption techniques, we observe that the long decay of the UP and LP features in strongly coupled cavity system corresponds to the time constant of triplets in the neat film (~ 25 μs) (**Figure S5** and **S6**), in support of our model.

The underlying justification for our model is that the presence of the polariton states only weakly perturbs the dynamics of the exciton reservoir. To further test this assertion, we use ultrafast time-resolved photoluminescence spectroscopy to directly monitor only the dynamics of emissive species (singlet excitons and cavity polaritons). These data allow us to simultaneously account for two important observations: 1) the dispersion in angle dependent measurements of the steady-state PL intensity indicates that emission occurs primarily through the LP and 2) the singlet exciton is populated for finite times (~ 50 ps) such that efficient SF is able to occur. At a collection angle of ~ 10°, the emission maximum near 665 nm decays with a rate constant of 2.0 x $10^{-2}$ $ps^{-1}$ (**Figure 4a**). This decay constant (total lifetime) represents the sum of all decay processes, including both SF and scattering to the LP. The correspondence to the SF rate constant $k_2$ (2.5 x $10^{-2}$ $ps^{-1}$) obtained from transient absorption implies weak scattering of singlet excitons to the LP followed by rapid (high yield) outcoupling at the LP lifetime (~0.05 $ps)^{-1}$. The measured total emission lifetime is a result of singlet excitons converting to the LP with a rate constant that is much less that the SF rate constant (~ 50 $ps)^{-1}$ but greater than the singlet radiative rate (~ 13,000

ps)[-1].[46,47] In other words, emission through the LP is the dominant radiative decay process, even if it represents an overall minority decay channel for photoexcited molecules. In agreement with a weak scattering process, we find that although the maxima of the 5 ps transient emission spectra shift according to the LP dispersion as a function of the collection angle (**Figure 4b**), the decay rate is angle independent (**Figure S8**), ruling out a direct population and decay of polaritons. The weak scattering rate is consistent with other recent measurements showing that differences in the density of states and delocalization between the exciton reservoir (*N*-1 modes) and the cavity polaritons (2 modes) limits the interconversion rate.[24,59]

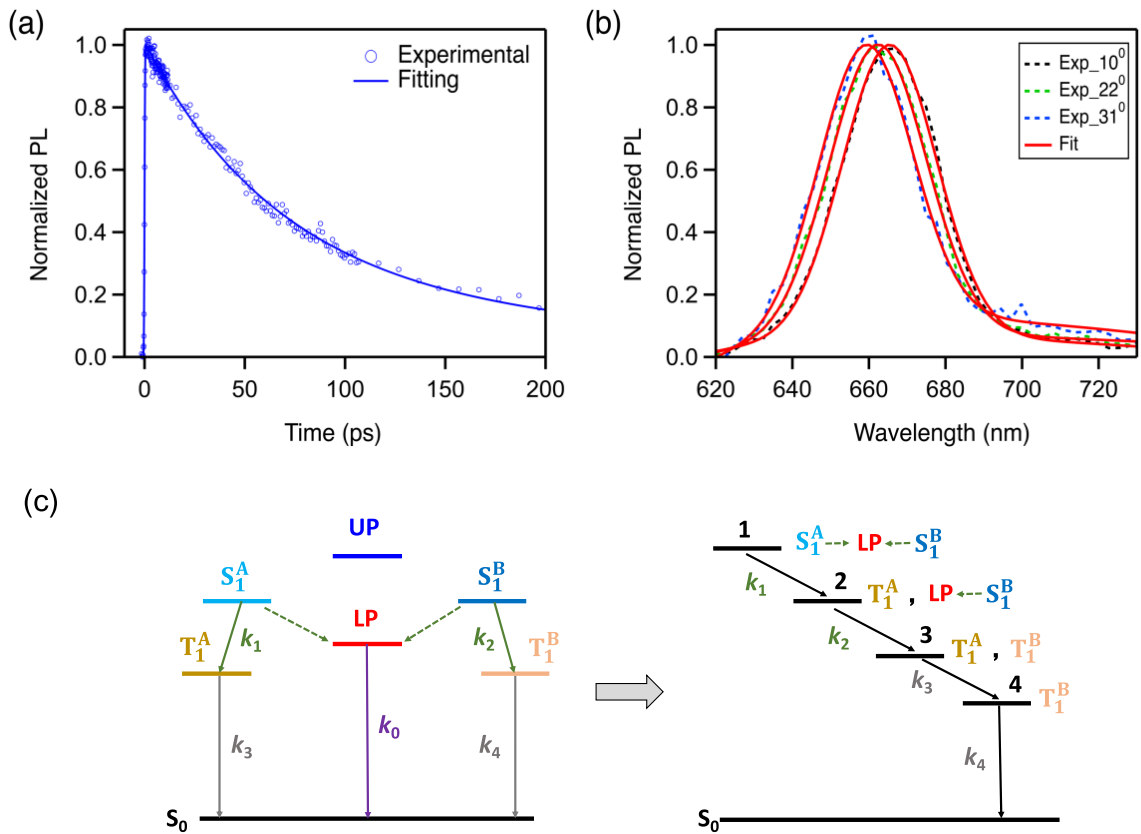

**Figure 4**. a) PL Intensity decay with the emission wavelength of 665 nm at 10º. b) Transient PL spectra of the strong coupled cavity system measured at several angles with a delay time of 5 ps. c) Illustration of the multilevel kinetic system for the strongly coupled cavity. The left illustration

includes two distinct exciton populations ($S_1^A$, $S_1^B$) that decay independently, primarily via singlet fission to triplet excitons ($T_1^A$, $T_1^B$) with a small contribution from scattering (dotted line) to the lower polariton (LP). Triplet states recombine on long times to the ground state ($S_0$), while LP population will rapidly radiatively decay. The right illustrates the sequential evolution in our simplified model, in which the individual species are a linear combination of the instantaneous exciton population ($S_1^A$, $S_1^B$, $T_1^A$, $T_1^B$) and the LP populated by weak scattering from $S_1^A$ and $S_1^B$. We note that the contribution of LP to the transient spectra is negligible due to the small instantaneous population ($< 0.1$ %).

These data allow us to build a full kinetic model for our strongly coupled cavity system (**Figure 4c**) in which the population dynamics of the organic film is only weakly perturbed by the presence of the cavity polaritons. The dynamical picture is similar to the neat (*half cavity*) films, with an inhomogeneous singlet and triplet population whose average behavior can be captured with two distinct sub-ensembles. As with the neat films, we can represent the dynamical picture using a branched model (**Figure 4c**, **left**) or with a sequential model (**Figure 4c**, **right**) that allows us to more readily identify and assign the contributing exciton states and the kinetic processes that connect them. The only modification to the neat film kinetic picture is the weak scattering of the singlet excitons into the LP followed by rapid emission, denoted as ($S_1^A \rightarrow LP \leftarrow S_1^B$) in the sequential model. Despite including it in our schematic, we note that the instantaneous population of the LP at finite times ($> 100$ fs) is $< 0.1$% as determined by a simple ratio of the rate constants for formation ($> 50$ ps) and decay ($< 50$ fs). The remaining exciton states and kinetic process are unchanged by strong coupling, with the $S_1^A$ singlet state undergoing singlet fission faster than the $S_1^B$, and the triplet state $T_1^A$ recombining faster than the $T_1^B$ triplet exciton.

**CONCLUSIONS**

Weak interactions between cavity polaritons and reservoir states requires a carefully considered approach to the design of polariton systems for modified chemical dynamics. First and foremost, this means that systems must be employed in which extremely long-lived excited states are present (e.g., triplet excitons) or systems in which light-matter interactions can be further enhanced. Promising systems fulfilling the latter requirement are open cavity (e.g., plasmonic) designs, in which the number of coupled molecules can be made to be extremely small, reducing the number of reservoir states. However, in typical organic microcavity systems with large ensemble of molecules, the primary role of the cavity polaritons will be as a sensitive reporter of changes in the molecular population dynamics. While weak coupling to the large reservoir of molecular excitons prohibits dramatic modifications of population dynamics in organic microcavity systems, these architectures may find utility in basic spectroscopic applications due to their high sensitivity to small modifications of the cavity refractive index.

**METHODS**

*Sample Fabrication*: A polymer matrix solution was prepared using polystyrene (Sigma-Aldrich, 182427-500G, Mw ~ 280,000) dissolved in toluene with a concentration of 20 mg/ml, and the singlet fission molecule was then added to the solution with a relative mass fraction of 30%. The near films for absorption and photoluminescence measurements were spin-cast on quartz-coated glass substrates.

For the microcavity fabrication, a 100 nm-thick Ag mirror was deposited by an e-beam evaporator in vacuum at $10^{-7}$ Torr, and the 160 nm TIPS-pentacene/PS cavity layer was then spin-cast on top of the Ag mirror. Microcavity fabrication was completed by depositing a second 30 nm-thick Ag layer on top of the organic film.

*Linearly Optical Spectroscopy*: Angle-resolved reflectivity and photoluminescence spectroscopy were measured using a home-built microphotoluminescence setup within the Fourier imaging configuration. For angle-resolved reflectivity measurements, a broadband of white light from tungsten halogen lamp was focused onto the sample by a 50x objective with a high numerical aperture (NA=0.8), and the reflected signal was collected by the same 50x objective, covering an angular range of ±53.1o, followed by the detection using the spectrometer (Princeton Instruments, Acton SpectraPro SP-2500) and charge-coupled device (CCD) camera (Princeton Instruments, PIX 1024B). For angle-resolved PL measurements, a fiber-coupled CW blue diode at 462 nm with a laser line filter was used to excite the sample, and the pump beam was focused onto the sample using the 50x objective (NA=0.8) with a spot size of 5 μm in diameter, and the emission signal was collected by the same 50x objective. A 600 nm long-pass filter was used to block the residual excitation beam.

*Transient Absorption Spectroscopy*: Transient absorption spectroscopy was performed using a commercial Ti:Sapphire laser system (1kHz repetition rate). A commercial optical parametric amplifier (TOPAS-C) was used to generate pump pulses with approximately 80 fs pulse widths. For femtosecond measurements, supercontinuum probe light was generated by focusing the 800 nm fundamental into a sapphire disc. The probe light was split into signal and reference beams, which were detected on a shot-by-shot basis by a fiber-coupled silicon visible diode array. The pump-probe delay was controlled by a mechanical delay stage. For longer delay times (ns - μs), a separate sub-ns supercontinuum laser is used to generate probe pulses that are electronically synchronized to the ultrafast laser (Ultrafast Systems EOS). All transient absorption spectroscopy was performed using the reflection configuration with a small incident angle (~ 10o) for the probe.

*Ultrafast Photoluminescence Spectroscopy*: Ultrafast photoluminescence was measured by the up-conversion technique. Briefly, the sample was excited with 600 nm, 100 fs laser pulse. Sum frequency generation was achieved by mixing the spontaneous emission with a "gate" pulse in a nonlinear crystal. The magnitude of the unconverted optical signal was proportional to the instantaneous photoluminescence intensity and was measured at intervals of delay between excitation and gate pulse.

*Global Analysis*: All transient absorption data was analyzed using the Wavemetrics Igor Pro software package. Briefly, singular value decomposition was used to reduce the size of the data set and identify the minimum number of species and time constants needed to represent the full data set. Global fitting using a parallel decay model (simple sum of exponentials) assuming first order decay of all species, followed by spectral decomposition (often referred to as decay associated spectra or DAS). While triplet annihilations process are fundamentally second order, the long time constant obtain in our system from the embedded TIPS-Pc is nearly identical to the solution phase triplet lifetime, implying that first order decay dominates over annihilation. Finally, a sequential model (1 → 2 → 3 → 4) was constructed from the parallel model using matrix diagonalization approaches, giving a new set of population dynamics and a new set of spectral decompositions that represent the "excited state spectra" (independent of time) of each step in the sequential model. As discussed in the text, we find that these sequential steps correspond to linear combinations of a set of exciton states that evolve over time.

## Corresponding Author

* Email: vmenon@ccny.cuny.edu, msfeir@gc.cuny.edu.

## Acknowledgements

The authors acknowledge support from the U.S. Department of Energy, Office of Basic Energy Sciences through Award No. DE-SC0017760. This research used resources of the Center for Functional Nanomaterials, which is a U.S. DOE Office of Science Facility, at Brookhaven National Laboratory under Contract No. DE-SC0012704. The authors also acknowledge the use of the Nanofabrication Facility at the Advanced Science Research Center of the City University of New York for the device fabrication.

**Author contributions**

M.Y.S. and V.M.M. conceived and supervised the project; B.L. and M.Y.S. designed and conducted the experiments, and performed data analysis. All authors contributed to write the manuscript and discuss the results.

**Additional information**

Supplementary Information is available